\documentclass[oldversion]{aa}

\usepackage{graphicx}

\def\smallsun{\hbox{$_\odot$}} 
\def\degr{\hbox{$^\circ$}}
\def\arcmin{\hbox{$^\prime$}}
\def\arcsec{\hbox{$^{\prime\prime}$}}
\def\cm3{cm$^{-3}$}
\def\ebv{$E_{\mathrm{B-V}}=$}

\begin{document}

%\thesaurus{09(09.01.1; 09.16.2; M1-42, 13.09.4)}

\title{Abundances of Planetary Nebula \object{M\,1-42}\thanks{Based on
    observations with the Spitzer Space Telescope, which is operated
    by the Jet Propulsion Laboratory, California Institute of
    Technology}}

\author{S.R.\,Pottasch\inst{1} \and J.\,Bernard-Salas\inst{2} \and
  T.L.\,Roellig\inst{3}}

\offprints{pottasch@astro.rug.nl} 

\institute{Kapteyn Astronomical Institute, P.O. Box 800, NL 9700 AV
  Groningen, the Netherlands \and Center for Radiophysics and Space
  Research, Cornell University, Ithaca, NY 14853 \and NASA Ames
  Research Center, MS 245-6, Moffett Field, CA 94035-1000}

\date{Received date /Accepted date}

\abstract{The spectra of the planetary nebula M\,1-42 is reanalysed
  using spectral measurements made in the mid-infrared with the
  Spitzer Space Telescope. The aim is to determine the chemical
  composition of this object.  We also make use of ISO, IUE and ground
  based spectra.  Abundances determined from the mid- and far-infrared
  lines, which are insensitive to electron temperature, are used as
  the basis for the determination of the composition, which are found
  to substantially differ from earlier results. High values of neon,
  argon and sulfur are found. They are higher than in other PN, with
  the exception of NGC\,6153, a nebula of very similar abundances. The
  high values of helium and nitrogen found indicate that the second
  dredge-up and hot bottom burning has occurred in the course of
  evolution and that the central star was originally more massive than
  4M\smallsun. The present temperature and luminosity of the central
  star is determined and at first sight may be inconsistent with such
  a high mass.
  }

\authorrunning{Pottasch et al.}

\titlerunning{Abundances in M\,1-42}  

\maketitle

\section{Introduction}

\object{M\,1-42} (\object{PN G2.7-04.8}) is a rather faint planetary nebula
(PN) with a high radial velocity which is located in the direction of
the galactic center. For this reason it is considered by Acker et al.
(\cite{acker}) to be a galactic bulge PN. It has been studied earlier
by Liu et al. (\cite{liu2}) who found that it has a very low Balmer
jump temperature and the abundances found from the measured
recombination lines are an order of magnitude higher than those found
from the collisional excited lines.

The purpose of this paper is to study the element abundances in this
nebula with the help of mid-infrared spectra taken by the Spitzer
Space Telescope (Werner et al. \cite{wer}). Abundances in these
nebulae have been studied earlier by Exter et al. (\cite{exter}),
Aller \& Keyes (\cite{aller}) and Ratag et al. (\cite{ratag}) using
only the visual spectrum as well as by Liu et al. (\cite{liu2}) who
also included ISO measurements. The use of the mid- and far-infrared
spectrum permits a more accurate determination of the abundances for
reasons that have been discussed in earlier papers (e.g. see Pottasch
\& Beintema \cite{pott1}; Pottasch et al. \cite{pott2}, \cite{pott4};
Bernard Salas et al. \cite{bernard}), and can be summarized as
follows.

The most important advantage is that the infrared lines originate from
very low energy levels and thus give an abundance which is not
sensitive to the temperature in the nebula, nor to possible
temperature fluctuations. Furthermore, when a line originating from a
high-lying energy level in the same ion is observed, it is possible to
determine an effective (Boltzmann) temperature at which the lines in
that particular ion are formed. When the effective temperature for
many ions can be determined, it may be possible to make a plot of
effective temperature against ionization potential, which can be used
to determine the effective temperature for ions for which only lines
originating from a high energy level are observed.  This may not work
if the same ion is present in a range of electron temperatures.

Use of the infrared spectra have further advantages. One of them is
that the number of observed ions used in the abundance analysis is
approximately doubled, which removes the need for using large
`Ionization Correction Factors' (ICFs), thus substantially lowering
the uncertainty in the abundance. A further advantage is that the
extinction in the infrared is almost negligible, eliminating the need
to include sometimes large correction factors. This is especially
important when the extinction is uncertain.  As mentioned above, Liu
et al. (\cite{liu2}) used the ISO mid- and far-infrared observations
in their analysis. We have used the IRS spectrograph of the Spitzer
Space Telescope to observe M1-42 in the same spectral region as the
ISO SWS measurements. We report different results, which are probably
due to two circumstances. First of all the SWS measurement was not
well centered on the nebula.  Together with the fact that the nebula
is rather large (about 11\arcsec~in diameter), this means that part of
the nebula may be missed in the ISO measurement (this is also true of
the IRS short wavelength slit). Furthermore the observed spectrum has
low intensity so that the errors in the ISO SWS intensities are high.

This paper is structured as follows. First the IRS spectrum of M1-42
is presented and discussed (in Sect. 2). Then the intrinsic H$\beta$
flux is determined using both the measurements of the infrared
hydrogen lines and the radio continuum flux density (Sect. 3). The
visible spectrum of the nebula is presented in Sect. 4 together with a
new reduction of the ultraviolet (IUE) spectrum. This is followed by a
discussion of the nebular electron temperature and density and the
chemical composition (Sect. 4). A comparison of the resultant
abundances with those made earlier is given in Sect. 5.  In this
section the relation between nebular abundances determined from
collisionally excited lines and from recombination lines is also
discussed, as well as the striking similarity of the abundances
obtained with those of NGC\,6153. In Sect. 6 discussion and concluding
remarks are given.

\section{The Infrared spectra}
\subsection{IRS observations of M\,1-42}

\begin{figure*}
  \centering
    \includegraphics[width=12cm,angle=90]{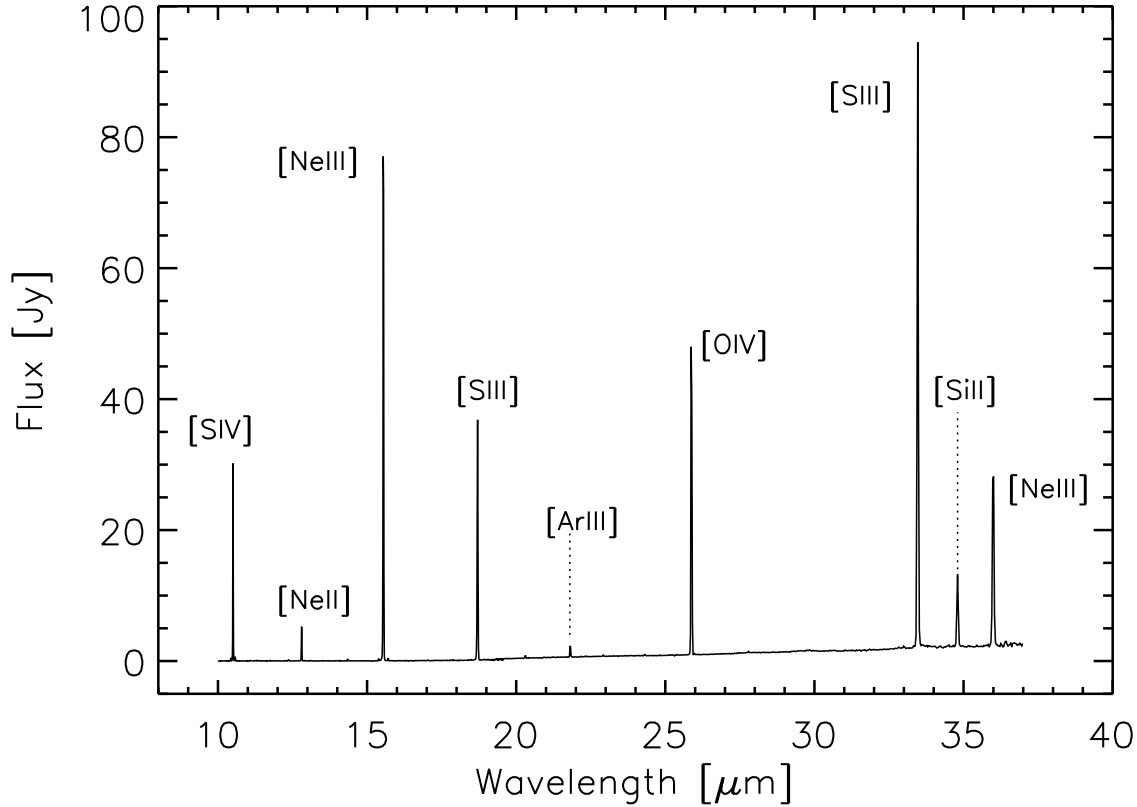}
    \caption{Observed Spitzer-IRS high-resolution spectrum of M\,1-42.
      The most prominent lines are labeled in the figure.}
  \label{m142_f}
\end{figure*}

The IRS spectrograph (Houck et al.\,\cite{hou}) on board the Spitzer
Space Telescope is similar to the ISO SWS spectrometer, but has higher
sensitivity. The IRS high resolution spectra have a spectral
resolution ($\lambda$/$\delta\lambda$) of about 600 which is less than
the ISO SWS by at least a factor of two. The aperture size is
comparable for the two instruments.  The SH has a slit size of
4.7\arcsec x 11.3\arcsec, while the LH is 11.1\arcsec x 22.3\arcsec,
the latter being close to the smaller SWS aperture.  The IRS high
resolution measures in two spectral ranges: the short high (SH) going
from 9.9$\mu$m to 19.6$\mu$m and the long high (LH) from 18.7$\mu$m to
37.2$\mu$m. The observed IRS spectrum of M\,1-42 is shown in
Figure~\ref{m142_f}.  Besides the usual strong lines of neon, sulfur,
oxygen and argon, many other lines have also been measured in the
spectrum,includung molecular hydrogen.  The ISO SWS has about the same
upper wavelength but extends on the short wavelength end to about
2.5$\mu$m.

The IRS measurement of M1-42 was centered at RA(2000)
18$^{h}$11$^{m}$04.87$^{s}$ and Dec(2000)
-28\degr58\arcmin59.4\arcsec.~This may be compared with the value
measured by Kerber et al. (\cite{kerber}) of RA(2000)
18$^{h}$11$^{m}$05.028$^{s}$ and Dec(2000)
-28\degr58\arcmin59.33\arcsec~from visual measurements. The ISO
measurement was made at RA(2000) 18$^{h}$11$^{m}$04.6$^{s}$ and
Dec(2000) -28\degr59\arcmin00.6\arcsec.  which is about
6.5\arcsec~from the position given by Kerber et al.  (\cite{kerber}) A
comparison of a photograph of the nebula with the apertures
superimposed indicates that the LH aperture measured the entire
nebula, while the SH aperture missed part of the nebula. A correction
was made for the missing emission by making use of the fact that the
two spectrographs had a wavelength region in common at about 19$\mu$m.
To make the continuum emission at this wavelength equal, the SH
emission had to be increased by a factor of 1.8. This factor is
approximately the ratio of the total area of the nebula to the area
measured measured in the SH slit (4.7x11.3\arcsec). This indicates
that the emission from the nebula is homogeneous. The measured
emission line intensities are given in Table 1, where the actual
measurements are given in the column labeled 'intensity'. In the
column giving the ratio of the intensity to H$\beta$ the SH intensity
includes the correction factor of 1.8; it also includes a small
correction for extinction using the extinction value given below and
the extinction curve given by Fluks et al. (\cite{fluks}). The
H$\beta$ intensity is discussed in Sect.3.

\begin{table}[htbp]
%Table 1\\
  \caption[]{The measured lines are
    given in Col.3. The last column gives the ratio of the line
    intensity to H$\beta$($=$100).}

%\begin{flushleft}
\begin{center}
\begin{tabular}{lrcc}
\hline
\hline
Ident. &  $\lambda$($\mu$m) & Intensity$^\dagger$ & I/H$\beta$$^\sharp$ \\
\hline

 ?                  & 9.98  & 6.52  & 0.96  \\
\ion{[S}{iv]}       & 10.51 & 1665  & 321   \\
\ion{H}{i} 9-7      & 11.30 & 4.27  & 0.806 \\
\ion{[Cl}{iv]}      & 11.76 & 4.50  & 0.844 \\
\ion{H$_2$}         & 12.28 & 3.53  & 0.65  \\
\ion{H}{i}7-6(11-8) & 12.37 & 10.2  & 1.89  \\
\ion{[Ne}{ii]}      & 12.81 & 235   & 43.5  \\
\ion{[Ar}{v]}       & 13.10 & 3.41  & 0.63  \\
\ion{[Ne}{v]}       & 14.31 & 2.78  & 0.51  \\
\ion{[Cl}{ii]}      & 14.36 & 7.68  & 1.41  \\
\ion{[Ne}{iii]}     & 15.55 & 2708  & 497   \\
\ion{H$_2$}         & 17.03 & 1.83  & 0.336 \\
\ion{[S}{iii]}      & 18.70 & 994   & 184   \\
\ion{[Cl}{iv]}      & 20.31 & 10.47 & 1.07  \\
\ion{[Ar}{iii]}     & 21.81 & 41.0  & 4.19  \\
\ion{[Fe}{iii]}     & 22.92 & 3.68  & 0.376 \\
\ion{[Ne}{v]}       & 24.31 & 3.89  & 0.396 \\
\ion{[O}{iv]}       & 25.87 & 900   & 91.5  \\
\ion{[Fe}{ii]}      & 25.98 & 2.51  & 0.255 \\
\ion{H}{i}9-8       & 27.78 & 2.66  & 0.27  \\
\ion{[S}{iii]}      & 33.46 & 1185  & 120   \\
\ion{[Si}{ii]}      & 34.80 & 158   & 16.0  \\
\ion{[Ne}{iii]}     & 35.99 & 424   & 42.8  \\

\hline

\end{tabular}
\end{center}

$^\dagger$The intensity is measured in units of
10$^{-14}$erg~cm$^{-2}$~s$^{-1}$.

$^\sharp$In the ratio of intensity to H$\beta$ the intensities below
19$\mu$m are multiplied by a factor 1.8 to correct for part of the
nebula being outside the diaphragm. A reddening correction is also
made but it is very small.

\end{table}

There are other ways of determining the factor which accounts for the
emission missing in the SH aperture. The method given above, that of
making use of the fact that the continua in the region of wavelength
overlap should be equal, is theoretically satisfying. But any pair of
lines can also be used as long as they originate from the same ion and
two other conditions exist: that each of the lines is measured in a
different aperture, and that the theoretical value of the ratio of the
two lines is only weakly dependent on the electron temperature and
density. Five candidate pairs of lines exist. The best is the ratio of
the \ion{[Ne}{iii]} lines at $\lambda$15.5 and $\lambda$36.0$\mu$m.
Theoretically the ratio of the intensities of these lines should be
about 11.8, with only a small dependance on nebular temperature and
density. Our measurements of these line pairs again leads to a value
of 1.8, the same correcting factor given as by the continua.  The
\ion{[Ne}{v]} and \ion{[Cl}{iv]} line pairs are only slightly more
dependant on the nebular properties and they are consistent with the
value of 1.8. Only the hydrogen lines do not give consistent values. A
value of 1.5 is obtained from the $\lambda$12.38$\mu$m line, if it is
a blend of the $\lambda$12.372$\mu$m and the $\lambda$12.387$\mu$m 7-6
and 11-8 transitions of hydrogen.  While this could be an error in the
intensity measured for this line, it is more likely there is another
line which contributes to the measured intensity.  The \ion{He}{i}
(7-6) transition is a likely possibility. The line at
$\lambda$11.30~$\mu$m also appears to be a blend of the hydrogen line
with a helium line and/or a line of \ion{[Cl}{i]}. It thus appears
that the total emission in the SH wavelength band is obtained by
increasing the measured intensities by a factor of 1.8$\pm$0.2.

\subsection{Comparison with the ISO SWS spectrum}

Comparing the present intensities with those found by ISO SWS (de
Graauw et al. 1999) measurements given by Liu et al. (\cite{liu2}) is
complicated by the mispointing of the ISO measurement. In addition,
the ISO SWS observation of M\,1-42 is noisy and is compromised by an
inconsistency between the up and down scans.  The up scan shows no
line emission (the spectrum seems pure noise) while the down scan does
have line emission.  The strongest line in the SWS spectrum
(\ion{S}{iii} line at 18.7~$\mu$m) in the down scan shows a strange
blue wing which is completely absent in the IRS spectrum. The nature
of these problems is not known (both scans should show the same
spectrum) and makes the SWS data of this object suspicious.  The two
lines measured by ISO SWS with the largest aperture (above 29$\mu$m
the ISO aperture was 20\arcsec~x 33\arcsec) have intensities within
20$\%$ of the values we have found, which, considering the rather low
quality of these ISO measurements, is reasonable agreement. The
\ion{[O}{iv]} at 25.9$\mu$m measured by ISO is more than a factor of 3
smaller than our value. At least part of this difference is probably
because the ISO aperture is now smaller than for the two longer
wavelength lines. The three ISO lines measured shortward of 19$\mu$m
(\ion{[S}{iv]}.  \ion{[Ne}{iii]} and \ion{[S}{iii]}) are about
30-40$\%$ less intense than Spitzer has measured. Because the Spitzer
measurements have to be increased by a factor of about 1.8, the ISO
measurements made with the 14\arcsec~x 27\arcsec aperture of these
three lines must be increased by about a factor of 3.

\subsection{ISO LWS spectrum}

The ISO LWS (Clegg et al. \cite{cle}) observation of M\,1-42 is used
here because it is made with a much larger aperture which has a
diameter of about 80\arcsec~ and includes the entire nebula. The on
and off source spectra are designated TDT 70302104 and 70302105
respectively.  Background emission is only important in the
\ion{[N}{ii]} and \ion{[C}{ii]} lines: the latter line is unusable for
this reason. The measured line intensities are given in Table 2.

\begin{table}[htbp]
%Table 2\\
  \caption[]{ISO LWS spectrum of \object{M\,1-42}. The measured lines are
    given in Col.3. The last column gives the ratio of the line
    intensity to H$\beta$($=$100). The intensities are measured in units of 
    10$^{-14}$ erg~cm$^{-2}$~s$^{-1}$. No reddening correction has been made.}

\begin{center}
\begin{tabular}{lrcc}
\hline
\hline
Ident. &  $\lambda$($\mu$m) & Intensity & I/H$\beta$ \\
\hline

\ion{[O}{iii]} & 51.83  & 3830  & 383  \\
\ion{[N}{iii]} & 57.32  & 2690  & 269  \\
\ion{[O}{i]}   & 63.19  & 318   & 31.8 \\
\ion{[O}{iii]} & 88.35  & 1800  & 180  \\
\ion{[N}{ii]}  & 121.9  & 59    & 5.9  \\

\hline

\end{tabular}
\end{center}

\end{table}

\section{Extinction}

There are several methods for obtaining the extinction: (1) comparison
of the radio emission with the H$\beta$ flux, (2) comparison of the
observed and theoretical Balmer decrement, (3) the dip at
$\lambda$2200\AA~ and, (4) photometry of the exciting star. Only the
first two methods are applicable here.

\subsection{The 6\,cm radio emission}

The 6\,cm flux density has been measured with the VLA by Zijlstra et
al. (\cite{zijlstra}). They find a value of 24 mJy. The nebula has
also been measured at 21\,cm by Condon \& Kaplan (\cite{condon}) who
find 28.6 mJy, which, if optically thin, would imply 24.2 mJy at
6\,cm. Using values of T$_e$ and helium abundance determined below
together with the equation quoted in Pottasch (\cite{potta}) this
implies an H$\beta$ flux of 8.4x10$^{-12}$ erg~cm$^{-2}$~s$^{-1}$.

\subsection{The H$\beta$ flux}

The H$\beta$ flux may also be found from the measured HI(9-8)
transition at 27.80\,$\mu$m given in Table 1. Using the theoretical
value of the ratio of this transition to H$\beta$ given by Hummer and
Storey (\cite{hummer}) at T$_e$ = 7500 K and a density of
10$^3$cm$^{-3}$ leads to H$\beta$ = 1.18x10$^{-11}$
erg~cm$^{-2}$~s$^{-1}$. This is 40\% higher than the value found from
the radio flux density. A value of 1.0x10$^{-11}$
erg~cm$^{-2}$~s$^{-1}$ will be used in the remainder of the paper. The
uncertainty is about 20\%.  The HI(7-6) transition could also have
been used to determine the H$\beta$ flux. There are two reasons for
not using this transition.  Firstly it is in the spectral region where
the slit did not cover the entire nebula. Secondly the line is blended
with another hydrogen line (HI(11-8)) and a helium line. Although
approximately the same H$\beta$ flux is obtained from this line, the
uncertainty is somewhat larger.

\subsection{The Balmer decrement extinction $E_{\mathrm{B-V}}$}

The extinction determined from the Balmer decrement differs somewhat
according to the author. Liu et al.(\cite{liu2}) give C=0.71, Aller \&
Keyes (\cite{aller}) give C=0.60, Exter et al.(\cite{exter}) find
C=0.53.

The measured value of the H$\beta$ flux (which is affected by
extinction) is 2.3x10$^{-12}$erg~cm$^{-2}$~s$^{-1}$ (Webster
\cite{webs}) which leads to a value of C=0.63 or
$E_{\mathrm{B-V}}$=0.43. This is in reasonably good agreement with the
values found from the Balmer decrement and will be used when necessary
in this paper.

\subsection{The visual spectrum}

The visual spectrum has been measured by several authors. The highest
resolution spectrum is by Liu et al. (\cite{liu2}), and reliable
spectra have also been reported by Exter et al. (\cite{exter}) and
Aller \& Keyes (\cite{aller}).  The measurements were made with slits
which cover a reasonably large area of the nebula and may be
considered as representative of the total nebula. The measurements of
Liu et al. (\cite{liu2}) are preferred because of the higher spectral
resolution used. Some weight is also given to the careful work of
Exter et al.  (\cite{exter}). The results are shown in Table 3 where
the Cols.3, 4 and 5 give the intensities measured by the various
authors relative to H$\beta$=100 for those lines which are of
interest. The line intensities have been corrected by the individual
authors for extinction. No attempt has been made to use a common
extinction correction because all give a correct Balmer decrement. In
the last column average values are given, with greater weight given to
the measurements of the first two authors.
     
\begin{table}[h]
%Table 3\\
\caption[]{Visual Spectrum of M\,1-42.}
\begin{center}
\begin{tabular}{llcccc}
\hline
\hline
\multicolumn{1}{c}{$\lambda$} & Ion & \multicolumn{3}{c}{Intensities}& Total\\ \cline{3-5}
\multicolumn{1}{c}{(\AA)}& & (1)$^\dagger$ & (2)$^\dagger$& (3)$^\dagger$ & Intens.\\
\hline

3726 & \ion{[O}{ii]}   & 38.5 & 34.4  & 51$^{*}$ & 36.0\\
3729 & \ion{[O}{ii]}   & 28.5 &  24.4 &          & 26.0 \\
3869 & \ion{[Ne}{iii]} & 67.4 & 67.2  & 53       & 67.3 \\
%4068 & \ion{[S}{ii]}  & 2.44 &  &  & 1.43 \\
%4076 & \ion{[S}{ii]}  & 0.88 &  &  & 0.51 \\
4102 & \ion{H$\delta$} & 25.9 & 25.4  & 26.9     & 25.8 \\
4267 & \ion{C}{ii}     & 2.88 & 2.55  & 2.24     & 2.55 \\
4340 & \ion{H$\gamma$} & 46.6 & 45.3  & 46.7     & 46.5 \\
4363 & \ion{[O}{iii]}  & 2.65 & 2.48  & 2.7      & 2.55 \\
4471 & \ion{He}{i}     & 7.03 & 7.28  & 8.0      & 7.3  \\
4686 & \ion{He}{ii}    & 11.2 & 13.5  & 11.7     & 12.5 \\
4711 & \ion{[Ar}{iv]}  & 2.07 &  1.81 & 2.7      & 2.0  \\
4724 & \ion{[Ne}{iv]}  &      &0.030  &          & 0.030\\
4725 & \ion{[Ne}{iv]}  &      &0.035  &          & 0.035\\
4740 & \ion{[Ar}{iv]}  & 1.22 &  1.38 & 1.7      & 1.38 \\
4861 & \ion{H$\beta$}  & 100  & 100   & 100      & 100  \\
5007 & \ion{[O}{iii]}  & 459  & 515   & 547      & 510  \\
5518 & \ion{[Cl}{iii]} & 0.42 & 0.648 & 1.33     & 0.65 \\
5538 & \ion{[Cl}{iii]} & 0.38 & 0.617 & 0.90     & 0.61 \\
5755 & \ion{[N}{ii]}   & 2.26 & 3.08  & 2.76     & 2.7  \\
5876 & \ion{He}{i}     & 16.9 & 23.8  & 22.6     & 22.6 \\
6312 & \ion{[S}{iii]}  & 1.02 & 1.44  & 1.39     & 1.42 \\
6563 & \ion{H$\alpha$} & 285  & 323   & 283      & 290  \\
6584 & \ion{[N}{ii]}   & 214  & 276   & 267      & 250  \\
6717 & \ion{[S}{ii]}   & 16.0 & 17.0  & 13.2     & 16.0 \\
6731 & \ion{[S}{ii]}   & 18.9 & 20.7  & 18.0     & 20.0 \\
7135 & \ion{[Ar}{iii]} & 21.9 & 18.4  & 18.6     & 18.4 \\
7320 & \ion{[O}{ii]}   &      & 2.37  &          & 2.37 \\
7330 & \ion{[O}{ii]}   &      & 2.04  &          & 2.04 \\

\hline
\end{tabular}
\end{center}

$^\dagger$(1) Exter et al. (\cite{exter}), (2) Liu et al. (\cite{liu2}),
(3) Aller \& Keyes (\cite{aller}).\\ 
 $^{*}$This is a blend of $\lambda$3726 and $\lambda$3729.
\end{table}

\subsection{The IUE ultraviolet spectrum }

Only a single low resolution IUE observation of this nebula exists. It
was taken with the large aperature (10\arcsec x 23\arcsec) with an
exposure time of 17\,100 seconds. In spite of this long exposure time
the spectrum is underexposed and only a few lines can be identified.
Liu et al. (\cite{liu2}) have discussed this measurement and
identified four lines in the spectrum. We have found the same four
lines in our analysis. The line intensities have been rederived for us
by W.A.  Feibelman and differ somewhat from those given by Liu et al.
(\cite{liu2}). They are shown in Table 3. Considering the bad quality
of the spectrum, the difference between our measurements and those of
Liu et al. (\cite{liu2}) probably reflect the uncertainty in the
intensities.

The extinction correction was made by assuming a theoretical ratio for
the \ion{He}{ii} line ratio $\lambda$1640/$\lambda$4686\,\AA~ at
T=10\,000 K and an N$_e$ of 10$^{3}$cm$^{-3}$. This leads to a value
of the $\lambda$1640\,\AA~ line as given in column (2) of Table 4.
Together with the measured intensity of the line and the assumption of
a normal extinction law (e.g. that of Fluks et al.\,\cite{fluks})
leads to a value of the extinction C which is 50\% higher than that
given above.  This was also noticed by Liu et al.\,(\cite{liu2}) who
ascribed this to an abnormal extinction law in this direction. While
this could be true there is no other evidence for it. The higher
intensity of the $\lambda$1640\AA~ line could be due to the poor
measurement. Below it is shown that the $\lambda$1663\AA~ line of
\ion{O}{iii]} leads to an improbably high electron temperature. The
conclusion, which we share with Liu et al.\,(\cite{liu2}), is that
this line is probably spurious and should be ignored. This is
indicative of the uncertainties in the UV intensities. If the
extinction correction is made relative to $\lambda$1640\,\AA~using
\ebv0.43, and the reddening curve of Fluks et al.(\cite{fluks}) we
obtain the results shown in the last column of Table 4.  Liu et
al.(\cite{liu2}) derive the intensities using an extinction correction
which is the same as ours in the visible and near UV but increases
more steeply toward the $\lambda$1640\AA~ line.  As a consequence the
\ion{C}{iii]} line at $\lambda$1909\AA~ has an intensity only half as
great as we have found. This increases the uncertainty of the carbon
abundance by a factor of two.

\begin{table}[htbp]
%Table 4\\
\caption[]{IUE Spectrum of M\,1-42.}  
\begin{center}
\begin{tabular}{llccc}
\hline
\hline
\multicolumn{1}{c}{$\lambda$} & Ion &\multicolumn{3}{c}{Intensities}\\
\cline{3-5}
\multicolumn{1}{c}{(\AA)}& & (1)$^\dagger$ & (2)$^\sharp$ & (3)(I/H$\beta$)$^\star$   \\
\hline

1640 & \ion{He}{ii}   & 4.91  & 8.1  & 81   \\
1663 & \ion{O}{iii]}  & 2.89  & 4.7  & 47   \\
1863 & \ion{Al}{iii]} & 1.71  & 2.8  & 28   \\
1909 & \ion{C}{iii]}  & 2.77  & 4.56 & 45.6 \\

\hline 
\end{tabular}
\end{center}
$^\dagger$Measured intensity from low resolution spectra in units of
 10$^{-14}$ erg cm$^{-2}$ s$^{-1}$. \\
$^\sharp$Intensity corrected for extinction in units of
 10$^{-12}$ erg cm$^{-2}$ s$^{-1}$. \\
$^\star$Normalized to H$\beta=$100.\\ 
\end{table}

\section{Chemical composition of the nebulae}

The method of analysis is the same as used in the papers cited in the
introduction. First the electron density $N_{\mathrm{e}}$ and
temperature $T_{\mathrm{e}}$ as function of the ionization potential
are determined. Then the ionic abundances are determined, using
density and temperature appropriate for the ion under consideration.
Then the element abundances are found for those elements in which a
sufficient number of ionic abundances have been derived.

\subsection{Electron density and distance}

The ions used to determine $N_{\mathrm{e}}$ are listed in the first
column of Table\,5. The ionization potential required to reach that
ionization stage, and the wavelengths of the lines used, are given in
Cols. 2 and 3 of the table. Note that the wavelength units are
\AA~when 4 ciphers are given and microns when 3 ciphers are shown. The
observed ratio of the lines is given in the fourth column; the
corresponding $N_{\mathrm{e}}$ is given in the fifth column. The
temperature used is discussed in the following section, but is
unimportant since these line ratios are essentially determined by the
density.

There is no indication that the electron density varies with
ionization potential in a systematic way. As already pointed out by
Liu et al.\,(\cite{liu2}) the \ion{[O}{iii]} lines always give a lower
density than the other lines. Ignoring these lines, the electron
density is 1400 cm$^{-3}$. The error is about 20\%.  It is interesting
to compare this value of the density with the rms density found from
the H$\beta$ line. This depends on the distance of the nebula which
isn't accurately known, and is estimated in the following paragraph as
5\,kpc. For this calculation we shall use a diameter of 10\arcsec.
Both of these quantities are uncertain. The H$\beta$ flux has been
given above and the electron temperature will be discussed below. We
obtain the uncertain rms values of 1200 cm$^{-3}$. Since both methods
are in fair agreement we will use their average, giving a density of
1300 cm$^{-3}$ in the further discussion of the abundances.  Notice
that the densities determined from the three optical line ratios are
the same as that determined from the infrared line ratios.

The distance to the nebula is estimated in the following way. The
position of the nebula is within a few degrees of the direction of the
galactic center.  Coupled with its high radial velocity it is almost
certain that it belongs to the group of galactic bulge planetary
nebulae. It is likely somewhat closer to us than the galactic center
for two reasons. First its diameter is larger than most galactic bulge
PNe even though it has a normal density. Second and perhaps more
important, the extinction to the nebula is not so large (see Sect.
3.3). Most galactic bulge PN have higher extinction. We conclude that
the PN belongs to the galactic bulge but is closer to us than many of
the galactic bulge nebulae. The actual value used (5\,kpc) only means
that the distance probably lies between 4 and 6\,kpc and it might be
even somewhat further away if it is in a region of reduced extinction.
Note that the statistical distances given by various authors (as
listed by Acker et al.,\cite{acker}) are somewhat smaller, ranging
from 2.6 to 4.2\,kpc and are also uncertain.

\begin{table}[t]
%Table 5\\
\caption[]{ Electron density indicators in M\,1-42.}
\begin{center}
\begin{tabular}{lcccc}
\hline
\hline
Ion &Ioniz. & Lines& Observed &N$_{\mathrm{e}}$ \\
&Pot.(eV) & Used  & Ratio & (cm$^{-3}$)\\
\hline
\ion{[S}{ii]}   & 10.4 & 6731/6716 & 1.25  & 1200 \\
\ion{[O}{ii]}   & 13.6 & 3626/3729 & 1.44  & 1600 \\
\ion{[S}{iii]}  & 23.3 & 33.5/18.7 & 0.65  & 1400 \\
\ion{[Cl}{iii]} & 23.8 & 5538/5518 & 0.924 & 1400 \\
\ion{[O}{iii]}  & 35.1 & 51.9/88.4 & 2.13  & 800  \\  
\hline
\end{tabular}
\end{center}

\end{table}

\subsection{Electron temperature}

A number of ions have lines originating from energy levels far enough
apart that their ratio is sensitive to the electron temperature. These
are listed in Table 6 which is arranged similarly to the previous
table. The electron temperature remains constant as a function of
ionization potential at a value of about 8\,000$\pm$1\,000.

%table 6

\begin{table}[t]
\caption[]{Electron temperature indicators in M\,1-42.}
%\begin{flushleft}
\begin{center}
\begin{tabular}{lcccc}
\hline
\hline
 Ion & Ioniz. & Lines& Observed & $T_{\mathrm{e}}$\\
 & Pot.(eV)& Used &Ratio  & (K) \\
\hline

\ion{[N}{ii]}   & 14.5 & 5755/6584 & 0.0108 & 8\,400 \\
\ion{[S}{iii]}  & 23.3 & 6312/18.7 & 0.0079 & 7\,500 \\
\ion{[Ar}{iii]} & 27.6 & 7136/21.8 & 4.39   & 6\,500\\
\ion{[O}{iii]}  & 35.1 & 4363/5007 & 0.0050 & 9\,300\\
\ion{[O}{iii]}  & 35.1 & 5007/51.8 & 2.83   & 7\,800\\
\ion{[Ne}{iii]} & 41.0 & 3869/15.5 & 0.136  & 7\,000\\

\hline
\end{tabular}
\end{center}
\end{table}

\subsection{Ionic and element abundances}

The ionic abundances have been determined using the following equation:

\begin{equation}
\frac{N_{\mathrm{ion}}}{N_{\mathrm{p}}}= \frac{I_{\mathrm{ion}}}{I_{\mathrm{H_{\beta}}
}} N_{\mathrm{e}}
\frac{\lambda_{\mathrm{ul}}}{\lambda_{\mathrm{H_{\beta}}}} \frac{\alpha_{\mathrm{H_{\beta}}}}{A_{\mathrm{ul}}}
\left( \frac{N_{\mathrm{u}}}{N_{\mathrm{ion}}} \right)^{-1} 
\label{eq_abun}
\end{equation}

where $I_{\mathrm{ion}}$/$I_{\mathrm{H_{\beta}}}$ is the measured
intensity of the ionic line compared to H$\beta$, $N_{\mathrm{p}}$ is
the density of ionized hydrogen, $\lambda_{\mathrm{ul}}$ is the
wavelength of this line, $\lambda_{\mathrm{H_\beta}}$ is the
wavelength of H$\beta$, ${\alpha_{\mathrm{H_\beta}}}$ is the effective
recombination coefficient for H$\beta$, $A_{\mathrm{ul}}$ is the
Einstein spontaneous transition rate for the line, and
$N_{\mathrm{u}}$/$N_{\mathrm{ion}}$ is the ratio of the population of
the level from which the line originates to the total population of
the ion. This ratio has been determined using a five level atom.

The results are given in Table 7, where the first column lists the ion
concerned, and the second column the line used for the abundance
determination. The third column gives the intensity of the line used
relative to H$\beta$=100. The fourth column gives the ionic abundances
assuming $T_{\mathrm{e}}$=8\,000 K, and the fifth column gives the
Ionization Correction Factor (ICF). For the first eight elements in
the table we judge that all the important ionization stages are
observed so that the ICF is unity (or almost unity). For both Fe and
Si a correction must be made for higher ionization stages than have
been observed. We have done this by comparing the distribution of
ionization stages of oxygen, neon and sulfur with what might be
expected for Fe and Si. These ionization correction factors are quite
uncertain. The element abundances, given in the last column, are
therefore probably well determined only for the first eight elements.
The carbon recombination line abundance is given at the end of the
table.  These abundances will be discussed in the next sections.

%Table 7\\
\begin{table}[htbp]
\caption[]
{Ionic concentrations and chemical abundances in \object{M\,1-42}.
Wavelength in Angstrom for all values of $\lambda$ above 1000, otherwise
in $\mu$m.}

%\begin{flushleft}
\begin{center}

\begin{tabular}{lccccc}
\hline
\hline
Ion & $\lambda$ &  I/H$\beta$$^\dagger$ & $N_{\mathrm{ion}}$/$N_{\mathrm{p}}$ 
&ICF & $N_{\mathrm{el.}}$/$N_{\mathrm{p}}$\\
\hline
He$^{+}$  & 5875 & 22.6  & 0.151    &     &          \\
He$^{++}$ & 4686 & 12.5  & 0.010    & 1   & 0.161    \\
C$^{++}$  & 1909 & 45.6  & 8.8(-4)  & 1.2 & 10.5(-4) \\
N$^{+}$   & 6584 & 250   & 9.43(-5) &     &          \\
N$^{+}$   & 122. & 5.9   & 1.5(-4)  &     &          \\
N$^{++}$  & 57   & 269   & 5.46(-4) & 1.1 & 7.5(-4)  \\
O$^{+}$   & 3727 & 36.0  & 8.5(-5)  &     &          \\
O$^{++}$  & 5007 & 510   & 4.42(-4) &     &          \\
O$^{++}$  & 51.8 & 383   & 7.25(-4) &     &          \\
O$^{++}$  & 88.3 & 180   & 9.5(-4)  &     &          \\
O$^{+3}$  & 25.8 & 91.5  & 2.4(-5)  & 1.0 & 8.3(-4)  \\
Ne$^{+}$  & 12.8 & 43.5  & 8.0(-5)  &     &          \\
Ne$^{++}$ & 15.5 & 497   & 3.6(-4)  &     &          \\
Ne$^{++}$ & 36.0 & 42.8  & 3.6(-4)  &     &          \\
Ne$^{++}$ & 3869 & 68.0  & 1.93(-4) &     &          \\
Ne$^{+4}$ & 24.3 & 0.396 & 4.3(-8)  & 1   & 4.4(-4)  \\  
S$^{+}$   & 6731 & 20.0  & 1.93(-6) &     &          \\
S$^{++}$  & 18.7 & 184   & 1.92(-5) &     &          \\
S$^{++}$  & 6312 & 1.44  & 9.53(-6) &     &          \\
S$^{+3}$  & 10.5 & 321   & 9.3(-6)  & 1.0 & 2.8(-5)  \\
Ar$^{++}$ & 21.8 & 4.19  & 7.1(-6)  &     &          \\
Ar$^{++}$ & 7135 & 18.4  & 3.14(-6) &     &          \\
Ar$^{+3}$ & 4740 & 1.38  & 1.07(-6) &     &          \\
Ar$^{+4}$ & 13.1 & 0.63  & 2.05(-8) & 1.1 & 9.0(-6)  \\
Cl$^{+}$  & 14.4 & 1.41  & 3.6(-7)  &     &          \\  
Cl$^{++}$ & 5538 & 0.61  & 1.96(-7) &     &          \\ 
Cl$^{+3}$ & 11.8 & 0.844 & 5.8(-8)  & 1.0 & 6.1(-7)  \\
Fe$^{+}$  & 26.0 & 0.255 & 6.7(-8)  &     &          \\
Fe$^{++}$ & 22.9 & 0.376 & 1.44(-7) & 2:  & 4.0(-7): \\ 
Si$^{+}$  & 34.8 & 0.16  & 4.8(-6)  & 2:  & 9.6(-6): \\ 
C$^{++}$  & 4267 & 2.5   & 1.9(-3)  &     &          \\
 
\hline
\end{tabular}
\end{center}
$^\dagger$Intensities given with respect to H$\beta=$100.\\
(:) indicates uncertain value\\
\end{table}

\section{Comparison with other abundance determinations}

Table 8 shows a comparison of our abundances with the most important
determinations in the past 20 years. There is marginal agreement,
usually to within a factor of two or three. It is rather surprising
that such differences exist, even for oxygen. We discuss below reasons
for this, which differ for different elements. A comparison is also
made in the table with the solar abundance (Asplund et al.
\cite{asplund}).  Note that for sulfur and chlorine more weight has
been given to the abundance determination in meteorites since this
determination is more accurate than for the sun itself. Neon and argon
abundances are taken from the references given in Pottasch and
Bernard-Salas (\cite{pbs}).

The helium abundance has been derived using the results of Benjamin et
al. (\cite{benjamin}). For recombination of singly ionized helium,
most weight is given to the $\lambda$ 5875\AA~line, because the
theoretical determination of this line is the most reliable.

%table 8
\begin{table}[htbp]
\caption[]{Comparison of abundances in \object{M\,1-42}.}
\begin{tabular}{lrrrrrr}
\hline
\hline
Elem.  & Present & Liu   & Exter & Aller  & Solar & N6153 \\ 
\hline  

He & 0.161  & 0.148 & 0.151 & 0.17 &  0.098 & 0.14  \\
C(-4)  & 10.5 & 0.63 &   &    &  2.5 & 6.8 \\
N(-4)  & 7.5  & 4.8  & 3.2  & 8.2 &  0.84 & 4.8 \\
O(-4)  & 8.3 & 4.3 & 3.0  & 4.7  &  4.6  & 8.3 \\
Ne(-4) & 4.4 & 1.3 & 1.2 &  1.35 & 1.2 & 3.1 \\
S(-5)  & 2.8 & 1.2 & 0.71  & 1.9 & 1.4 & 1.9 \\   
Ar(-6) & 9.0 & 3.7 & 3.4 &  4.6 &   4.2 & 8.5 \\
Cl(-7) & 6.1 & 1.9 &     & 7.1 & 3.2  & 5.6 \\

\hline  

\end{tabular}

References: Liu: Liu et al. (\cite{liu2}), Exter: Exter et
al. (\cite{exter}), Aller: Aller \& Keyes (\cite{aller}),  Solar:
Asplund et al. (\cite{asplund}), N6153: Pottasch et al. (\cite{6153}).
\end{table}

\subsection{Errors}

Here we will discuss the reasons for the abundance differences between
what we have obtained and what has been found by other authors. This
is interesting because we have used essentially the same visual
measurements as these groups, and the comparison provides some insight
into the reasons for the differences. From this it can be seen that
the errors occur not only as a result of measurement errors, but they
occur in the interpretation of the measurements: in the electron
temperature which is derived and in the corrections made for unseen
stages of ionization. It is even possible that errors in collisional
cross-sections play a role when different lines are used to analyse a
given ion. The ultraviolet measurements are somewhat different but
affect only the carbon abundance.

The error of measurement for the IRS intensities given in Table 1 is
small, probably not more than 5\%. A correction for missing emission
in the short wavelength band is probably of the same order. The
uncertainty in the collisional strengths introduce an error of 15-20\%
so that the total error for the ions of neon, sulfur and argon
determined here with the IRS measurements is less than 25\%. This will
also be true for the total abundances of these elements because the
ICF is about unity. The error for the nitrogen and oxygen abundance is
somewhat higher because the visual measurements are less certain
(compare the different intensities found by different authors shown in
Table 3). In addition the temperature is more important for these
lines and the total errors may be twice as large. The largest error is
for the carbon abundance which has already been discussed.

Bearing this in mind, it can be seen in Table 8 that our results
sometimes differ by a large factor with previous determinations. Here
we discuss the reasons for this for individual elements.

The differences with Liu et al. (\cite{liu2}) arise from two factors.
First of all these authors use a higher electron temperature, derived
from the \ion{[O}{iii]} $\lambda$4363/$\lambda$5007 line ratio. While
we find the same temperature from this ratio, it is inconsistent with
the lower temperature found from other line ratios (see Table 6) and
with the \ion{[O}{iii]} $\lambda$5007/$\lambda$51.8 line ratio
temperature. Secondly these authors use suspiciously low line
intensities in the SWS infrared region of the spectrum.  These effects
both work in the same direction to decrease the abundance.  A case in
point is the Ne$^{++}$ abundance for which we find an abundance of a
factor of 3 higher from the $\lambda$15.5$\mu$m line. The abundance
differences with Exter et al. (\cite{exter}) arise from the same
temperature difference plus the fact that they did not have the
infrared spectrum at their disposal and have to use ICFs. Aller \&
Keyes (\cite{aller}) used a temperature more similar to the one we
used and therefore the differences are not as marked.

The problem of the temperature is not yet completely solved. This
nebula does not show a marked temperature gradient as has been found
in some nebulae. It is possible that temperature fluctuations do exist
in this (and possibly other) nebula. For this reason it is desireable
to make use of the infrared spectrum whereever possible, since the
abundances derived from it do not have an important temperature
dependence. The temperature uncertainty has the largest effect for
carbon for which we find a value which is an order of magnitude higher
than what Liu et al.(\cite{liu2}) find. Interestingly, this carbon
abundance is almost equal to that found for M\,1-42 from the
recombination line of carbon, which we discuss in Sect.5.3.

\subsection{Recombination line abundances}

Recombination lines of oxygen, carbon, nitrogen, neon and magnesium
have been measured in this nebula and in several other PNe in recent
years. Abundances determinations using these recombination lines often
give higher abundances of the above elements (with the exception of
magnesium) than the determination using the collisional lines (e.g.
see Liu et al.,\cite{liu2}). The ratio of these two determinations
varies from one nebula to the other, from about 10 down to about 1.
The highest value is found in M\,1-42. The reason for this is at
present unknown. The only clues are that the recombination lines of
oxygen seem to be formed at a position different from the collisional
lines of oxygen and closer to the center of the nebula.
\footnote{Barlow has given a summary of these measurements in a recent
  workshop held in Beijing; they may be found on the website
http://ast.pku.edu.cn/$\sim$xs2007/
} 

Also they are likely formed in a region of considerably lower electron
temperature since these regions do not produce collisional line
emission. At the same time a lower temperature can explain the lower
electron temperature found from the Balmer jump. It has been suggested
that the recombination lines are produced in small inclusions of high
density and low temperature (see the workshop website in footnote).
The existence of such inclusions is a real possibility but still
unproved. If such regions exist they could affect our results because
the temperature, while being low, might still be high enough that
collisionally excited infrared lines could be formed there. No
evidence for this is seen in our measurements.  The abundances of
those ions which can be measured both from infrared lines and from
optical lines are very similar. At the same time the densities found
from infrared line ratios (\ion{[S}{iii]} and \ion{[O}{iii]}) are
essentilly the same as the densities determined from optical line
ratios and do not suggest that they are formed in high density regions
(see Table 5). Thus it seems unlikely that a great deal of the
infrared emission from these lines originates in high density
inclusions.

A strong indication that the abundances we have found are
representative of the majority of the material in the nebula is the
fact that they are in agreement with abundances found in all other PNe
we have studied, when the galactic gradient is taken into account (see
Pottasch \& Bernard-Salas\,\cite{pbs}) for oxygen, sulfur, neon and
argon which are not formed in the course of evolution. This is
unlikely to be a random occurence and thus lends support to the
reality of the abundances.

\subsection{Recombination line abundances of carbon}

The C$^{++}$ population can be obtained from the recombination line
$\lambda$4267\AA~as well as from the collisionally excited line at
$\lambda$1909\AA. The advantage of the recombination line is that it
is not sensitive to the electron temperature, and is in the visual
spectrum as well. It has the disadvantage that it is quite faint and
thus difficult to measure accurately. It has been used for at least 30
years, and has been found to sometimes give higher C$^{++}$
populations than the collisional line. The reason for this is not
clear (e.g. see the discussion of Liu et al. \cite{liu1} and the
discussion in the previous subsection). We have redetermined the
C$^{++}$ abundance using this line and the result is shown in the last
line of Table 7.  This is essentially the same value as is given by
Liu et al. (\cite{liu2}) using the recombination line.  While it is
still a factor of 2 higher than that determined from the
$\lambda$1909\AA~ line the errors are quite large. It is therefore
possible that both determinations give a similar result. While this
may be true of carbon, the recombination lines of oxygen, nitrogen and
neon give a much higher abundance for these elements than we have
found (Liu et al., \cite{liu2}).

\subsection{Comparison with NGC\,6153} 
 
The abundances in NGC\,6153 are shown in the last column of Table 8.
There is a remarkable similarity with those of M\,1-42. Although the
abundances in M\,1-42 may on average be slightly higher, the
uncertainty in the individual elements is at least 20\% to 30\% and
considerably higher for carbon. There are more similarities. First,
the Balmer jump temperature for both nebulae is extremely low, of the
order of 4000~K. Second, the abundances found from the recombination
lines is in both nebulae an order of magnitude higher than that found
from the lines excited by collisions. Third, the morphology of both
nebulae is similar in that no structure is seen in either nebula, yet
neither is clearly round or elliptical.

The electron temperatures of both nebulae are very similar. At first
sight this is expected since the abundances are almost the same. But
even the details of the various temperature determinations are
similar. For example, the temperature derived from the \ion{[O}{iii]}
$\lambda$4363\AA/5007\AA~ ratio is higher than the average in both
nebulae while that derived from the \ion{[O}{iii]} $\lambda$
5007\AA/51.8$\mu$m ratio is lower than the average. Furthermore the
temperatures derived from the \ion{[Ar}{iii]} and \ion{[S}{iii]} line
ratios are lower than the average in both nebulae. It seems to us that
these similarities are real and indicate that the same physical
situation exists in both nebulae, and that this physical situation is
the cause of the various, at first sight anomolous, measurements found
in these nebulae. These are thus clues as to the physical situation
existing in these nebulae.
 
\section{The central star}
\subsection{Stellar temperature}

Enough information is available to compute both the Zanstra
temperature and the Energy Balance temperature of the central star.
The Zanstra temperature requires the knowledge of the stellar apparent
magnitude, the extinction and the H$\beta$ flux. The last two
quantities have already been given in Sect.\,3. The apparent magnitude
is listed by Acker et al,(\cite{acker}) as V$=$17.4. Assuming that the
star radiates as a blackbody the hydrogen Zanstra temperature is
$T_{\mathrm{z}}$(H)= 63\,000 K and the ionized helium Zanstra
temperature is 81\,000 K. The Energy Balance temperature requires the
knowledge of the ratio of the 'forbidden' line intensities to
H$\beta$. This value is found by summing the intensities given above,
and is about 35 after making a correction for unmeasured lines using
the table of Preite-Martinez \& Pottasch\,(\cite{pmp}). This is a
rather uncertain number. To convert this value to a stellar
temperature, the formulation of Preite-Martinez \&
Pottasch\,(\cite{pmp}) is used, assuming blackbody radiation from the
central star. The value of Case II (the nebula is optically thin for
radiation which will ionize hydrogen but optically thick for ionized
helium radiation) for the energy balance temperature is
($T_{\mathrm{EB}}$)= 100\,000 K. If a model atmosphere had been used
instead of a blackbody, the energy balance temperature would be lower.
The low value of the hydrogen Zanstra temperature
($T_{\mathrm{z}}$(H)) may be due to the nebula being optically thin to
radiation ionizing hydrogen. An average stellar temperature of about
85\,000 K is a reasonable first approximation. This is indicated by
the rather good agreement of the various methods.

\subsection{Radius and luminosity}

The stellar radius and luminosity are dependent on the distance of the
nebulae which is difficult to obtain accurately. In Sect.\,4.1 we
estimated that the nebula is on the near side of the galactic center
and its distance hss an uncertain value of d=5 kpc. As a galactic
bulge nebula it could have a distance of 8 kpc. These distances,
combined with the magnitude of the star, lead to the stellar radius of
0.171 R\smallsun~and 0.438 R\smallsun. Once the radius is known the
gravity may be computed assuming that the star has a mass of 0.6
M\smallsun.  The value is g=5.6x10$^5$ and 2.2x10$^5$ cm s$^{-2}$.
Knowing the temperature and radius of the star the luminosity may be
found: L/L\smallsun=1400 at 5 kpc and 3500 at the less likely distance
of 8 kpc.

\section{Evolutionary considerations and conclusions}

The abundances in M\,1-42 are, at first sight, rather high. This
probably is related to the fact that it is in the galactic bulge. We
can use the PN abundances and abundance gradient given by Pottasch and
Bernard-Salas\,(\cite{pbs}) to estimate the abundance expected at 3
kpc from the galactic center, which would be the position of M\,1-42
if it were at a distance of 5 kpc from the sun. Their gradients lead
us to expect the following average abundances at this position:
O/H=12.6x10$^{-4}$,
Ne/H=3.2x10$^{-4}$,
S/H=2.1x10$^{-5}$,
Ar/H=8.0x10$^{-6}$.
These abundances are all within 50\% of the values found for M\,1-42.
From this we conclude that M\,1-42 has the abundances expected from a
PN at about this distance from the galactic center, although the
oxygen abundance is somewhat low compared to neon, sulfur and argon.
The helium abundance is high; this usually occurs in nebulae in which
hot bottom burning has taken place. But most models do not predict
values of He/H higher than 0.15 at any stage of their development. The
high nitrogen/oxygen ratio indicates that hot bottom burning has
occurred even though the ratio may be slightly lower than unity.
According to Karakas\,(\cite{karakas}) this will occur in stars of
masses greater than about 5M\smallsun. The same models show that the
carbon abundance increases relative to oxygen above 2.5M\smallsun; the
carbon/oxygen ratio becomes greater than unity above 3M\smallsun,
reaches a maximum at 3.5M\smallsun~and begins to decline at higher
masses. Carbon is expected to be a factor of at least 2 lower than
oxygen when nitrogen is as abundant as oxygen.  Because the carbon
abundance in M\,1-42 is uncertain, we do not put much weight on this
discrepancy. Thus it seems likely that M\,1-42 originated with a mass
of slightly more than 5M\smallsun. If this is so the rather low
luminosity of the central star of this nebula, although quite
uncertain, appears at first sight to be inconsistent with this high
mass. This is because calculations of the evolution of temperature and
luminosity of stars of different initial masses (e.g. Blocker et al.\,
\cite{blocker}) lead to the conclusion that a star of initial mass of
5M\smallsun~will evolve into a central star of core mass of
0.7M\smallsun~with a luminosity of log\,L=4. This luminosity is
considerably higher than we have found. The only way to reconcile
these seemingly discrepant results would be if the 5M\smallsun~star
evolved into a star of 0.5M\smallsun~core mass, due perhaps to a
higher mass loss after the nuclear processes have taken place.

\section{Acknowledgement}

This work is based on observations made with the Spitzer Space
Telescope, which is operated by the Jet Propulsion Laboratory,
California Institute of Technology under NASA contract 1407. Support
for this work was provided by NASA through Contract Number 1257184
issued by JPL/Caltech.

\end{document}